\documentstyle[aps,epsf,rotate]{revtex}

\draft

\begin{document}
\title{Shape of Crossover Between Mean-Field and Asymptotic Critical Behavior
in a Three-Dimensional Ising Lattice}
\author{M. A. Anisimov$^{1,2}$, E. Luijten$^{3,4}$, V. A. Agayan$^{1}$,
J. V. Sengers%
$^{1,2,}\thanks{%
Corresponding author. E-mail: sengers@ipst.umd.edu}$, and K. Binder$^{4}$}
\address{$^{1}$Institute for Physical Science and Technology, \\
University of Maryland, College Park, MD 20742, USA \\
$^{2}$Department of Chemical Engineering, University of Maryland, College
Park, MD 20742, USA \\
$^{3}$ Max-Planck-Institut f\"{u}r Polymerforschung, Postfach 3148, D-55021,
Mainz, Germany \\
$^{4}$Institut f\"{u}r Physik, WA 331, Johannes Gutenberg-Universit\"{a}t,
D-55099 Mainz, Germany} \maketitle

\begin{abstract}
  Recent numerical studies of the susceptibility of the three-dimensional Ising
  model with various interaction ranges have been analyzed with a crossover
  model based on renormalization-group matching theory. It is shown that the
  model yields an accurate description of the crossover function for the
  susceptibility.
\end{abstract}

\vspace{\baselineskip}

Recently, an accurate numerical study of the crossover from asymptotic
(Ising-like) critical behavior to classical (mean-field) behavior has been
performed both for two-dimensional \cite{LB:96,LB:97} and three-dimensional
\cite{LB:99} Ising systems in zero field on either side of the critical
temperature with a variety of interaction ranges. It is the objective of the
present work to analyze these numerical results within the framework of a
crossover theory that is based on renormalization-group matching and that
has already successfully been applied to the description of crossover in
several experimental systems~\cite{AS:95,MH:97}.

Qualitatively, the crossover is ruled by the parameter $t/G$ where $t=(T-T_{%
{\rm c}})/T_{{\rm c}}$ is the reduced temperature distance to the critical
temperature $T_{{\rm c}}$ and $G$ the Ginzburg number \cite{PP:79}. The
Ginzburg number depends on the normalized interaction range $R$ as
\begin{equation}
G=G_{0}R^{-2d/(4-d)} \;,
\end{equation}
where $d$ is the dimensionality of space and $G_{0}$ a constant. Hence, for $%
d=3$ the crossover occurs as a function of $tR^{6}$. Asymptotic critical
behavior takes place for $tR^{6}\ll 1$ and classical behavior is expected
for $tR^{6}\gg 1$. In real fluids the crossover is never completed in the
critical domain (where $t\ll 1$), since the range of interaction is of the
same order of magnitude as the distance between molecules ($R\simeq 1$) \cite
{AS:95}. A new Monte-Carlo algorithm, developed by Luijten and Bl\"ote \cite
{LB:95}, offers the advantage that the ratio $t/G$ can be tuned over more
than eight orders of magnitude allowing one to cover the full crossover
region in three-dimensional spin models~\cite{LB:99}.

A sensitive description of crossover behavior is obtained from an analysis
of the effective critical exponent of the susceptibility (the third
derivative of the free energy), defined as
\begin{equation}
\gamma _{{\rm eff}}^{\pm }\equiv -d\ln \hat{\chi}/d\ln |t|\;,  \label{Gammae}
\end{equation}
where the scaled susceptibility $\hat{\chi}=k_{B}T_{{\rm c}}(R)(\partial
m/\partial h)_{T}$, $k_{{\rm B}}$ the Boltzmann constant, $m$ the order
parameter, $h$ the ordering field, and where the ``$+$'' sign applies for $%
T>T_{{\rm c}}$, and the ``$-$'' sign for $T<T_{{\rm c}}$. As is seen from
Figs.~1 and~2, the variation of $\gamma _{{\rm eff}}^{\pm }$ reproduces the
Ising asymptotic critical behavior ($\gamma _{{\rm eff}}^{\pm }\simeq 1.24$)
at $tR^{6}\ll 1$ as well as the mean-field asymptote ($\gamma _{{\rm eff}%
}^{\pm }=\gamma _{{\rm MF}}=1$) at $tR^{6}\gg 1$. Apparently, all data would
seem to collapse onto a universal function of the reduced variable $tR^{6}$
as predicted by a field-theoretical treatment \cite{BB:84,BB:85} and by the $%
\varepsilon $-expansion \cite{BK:92}. However, as was noted in Ref.\ \cite
{LB:99}, a more careful look at the data reveals a remarkable discrepancy
between the theoretical calculations \cite{BB:84,BB:85,BK:92,SF:80,F:86}\
and the simulation results. Namely, the shape of the crossover is sharper
than predicted by the theory \cite{SF:80,F:86}, especially for short ranges
of interaction. We will show that this discrepancy is related to the
findings of Refs.~\cite{AS:95,MH:97}, where it was shown that there is a
fundamental problem in describing the crossover of $\gamma _{{\rm eff}}^\pm$
by a universal function which contains only a single crossover parameter $%
G\propto R^{-6}$.

In zero-ordering field above $T_{{\rm c}}$ the susceptibility asymptotically
close to the critical point behaves as
\begin{equation}
\chi =\Gamma _{0}t^{-\gamma }(1+\Gamma _{1}t^{\Delta _{{\rm s}}}+\Gamma
_{2}t^{2\Delta _{{\rm s}}}+a_{1}t+\ldots )\;,  \label{wegner}
\end{equation}
where $\gamma =1.239\pm 0.002$ (see, e.g., Refs.~\cite{SL:86,LF:89} and
references therein) and $\Delta _{{\rm s}}=0.504\pm 0.008$ \cite{GZJ:98} are
universal Ising critical exponents, and where $\Gamma _{0}$, $\Gamma _{1}$, $%
\Gamma _{2}$, and $a_{1}$ are system-dependent amplitudes. Expansion (\ref
{wegner}) is called the Wegner series \cite{We:72}.

In a universal single-parameter crossover theory \cite{BB:84,BB:85,BK:92},
the Ginzburg number is responsible both for the range of validity of the
mean-field approximation and for the convergence of the Wegner series (\ref
{wegner}). However, it is known \cite{NA:85,LF:90,TC:91}, that the sign of
the first Wegner correction amplitude $\Gamma _{1}$ depends on the
difference $u-u^{\ast }$, where $u$ is the scaled coupling constant and $%
u^{\ast }=0.472$ is the universal coupling constant at the Ising fixed
point~ \cite{GZ:87}. Moreover, Liu and Fisher \cite{LF:90} concluded that
the three-dimensional nearest-neighbor Ising model has a negative leading
Wegner correction amplitude $\Gamma _{1}$, so that $\gamma _{{\rm eff}}^{\pm
}$ asymptotically approaches $\gamma \simeq 1.24$ from above. Therefore,
since the coupling constant itself depends on the interaction range, the
shape of $\gamma _{{\rm eff}}^{\pm }$ cannot be represented by a universal
function of the Ginzburg number, since $G$ is not proportional to the
difference $u-u^{\ast }$.

In this paper we therefore present an analysis of the numerical data for $%
\gamma _{{\rm eff}}^{\pm }$ \cite{LB:99} in terms of a crossover model based
on renormalization-group matching for the free-energy density \cite
{NA:85,TC:91,AT:92}. This model contains two crossover parameters $\bar{u}%
=u/u^{\ast }$ and $\Lambda $ (a dimensionless cutoff wave number), and two
rescaled amplitudes $c_{t}$ and $c_{\rho }$ related to the coefficients of
the local density of the classical Landau--Ginzburg free energy $\Delta A$:
\begin{eqnarray}
\frac{v_{0}}{k_{{\rm B}}T}\frac{d(\Delta A)}{dV} &=&\frac{1}{2}a_{0}\tau
\varphi ^{2}+\frac{1}{4!}u_{0}\varphi ^{4}+\frac{1}{2}c_{0}(\nabla \varphi
)^{2}  \nonumber \\
&=&\frac{1}{2}c_{t}\tau M^{2}+\frac{1}{4!}u^{\ast }\bar{u}\Lambda M^{4}+%
\frac{1}{2}(\tilde{\nabla}M)^{2}\;,
\end{eqnarray}
with $\tau =(T-T_{{\rm c}})/T$, $M=c_{\rho }\varphi =\left(
a_{0}/c_{t}\right) ^{1/2}\varphi$, $a_{0}=c_{\rho}^{2}c_{t}$, $u_{0}=u^{\ast
}\bar{u}\Lambda c_{\rho }^{4}$, $c_{0}=c_{\rho }^{2}v_{0}^{2/3}$, and $%
\tilde{\nabla}=v_{0}^{1/3}\nabla $. The average molecular volume $v_{0}$ and
the prefactor $v_{0}/k_{{\rm B}}T$ are introduced to make the free-energy
density and all the coefficients dimensionless. The inverse crossover
susceptibility $\chi ^{-1}=\left( \partial ^{2}\Delta \tilde{A}/\partial
M^{2}\right) _{\tau }$, where $\Delta \tilde{A}$ is the crossover
(renormalized) free-energy density, in zero field above $T_{{\rm c}}$ reads
\cite{AS:95}
\begin{equation}
\chi ^{-1}=c_{\rho }^{2}c_{t}\tau Y^{(\gamma -1)/\Delta _{{\rm s}}}(1+y)
\label{chi}
\end{equation}
with
\begin{equation}
y=\frac{u^{\ast }\nu }{2\Delta _{{\rm s}}}\left\{ 2\left( \frac{\kappa }{%
\Lambda }\right) ^{2}\left[ 1+\left( \frac{\Lambda }{\kappa }\right) ^{2}%
\right] \left[ \frac{\nu }{\Delta _{{\rm s}}}+\frac{(1-\bar{u})Y}{1-(1-\bar{u%
})Y}\right] -\frac{2\nu -1}{\Delta _{{\rm s}}}\right\} ^{-1}\;,
\end{equation}
where $\nu \simeq 0.630$ \cite{GZJ:98,ic3d} is the critical exponent of the
asymptotic power law for the correlation length $\xi $ \cite{AS:95}. Note
that $\chi ^{-1}=(T_{{\rm c}}/T)\hat{\chi}^{-1}$ and the relation between $%
\gamma _{{\rm eff}}\equiv -d\ln \chi /d\ln |\tau |$ and $\gamma _{{\rm eff}%
}^{\pm }$, given by Eq.\ (\ref{Gammae}), is $\gamma _{{\rm eff}}^{\pm
}=\gamma _{{\rm eff}}+(1-\gamma _{{\rm eff}})\tau $, both above and below
the critical temperature. The crossover function $Y$ is defined by
\begin{equation}
1-(1-\bar{u})Y=\bar{u}\left[ 1+\left( \frac{\Lambda }{\kappa }\right) ^{2}%
\right] ^{1/2}Y^{\nu /\Delta _{{\rm s}}}  \label{YY}
\end{equation}
and is to be found numerically. The parameter $\kappa $ in Eq.\ (\ref{YY})
is inversely proportional to the fluctuation-induced portion of the
correlation length and serves as a measure of the distance to the critical
point. In zero field above $T_{{\rm c}}$ the expression for $\kappa ^{2}$
reads:
\begin{equation}
\kappa ^{2}=c_{t}\frac{T}{T_{{\rm c}}}\tau Y^{(2\nu -1)/\Delta _{{\rm s}%
}}=c_{t}tY^{(2\nu -1)/\Delta _{{\rm s}}}\;.  \label{kappa}
\end{equation}
We modified the original expression for $\kappa ^{2}$, given by Eq. (3) in
\cite{AS:95}, by introducing the non-asymptotic factor $T/T_{{\rm c}}$ in
Eq. (8) so that $\kappa ^{2}$ becomes infinite at $T\rightarrow \infty $
\cite{KS:??}. Asymptotically close to the critical point ($\Lambda /\kappa
\gg 1$), the following expression is obtained for the first correction
amplitude $\Gamma _{1}$ in Eq.\ (\ref{wegner}):
\begin{equation}
\Gamma _{1}=g_{1}\left( \frac{\sqrt{c_{t}}}{\bar{u}\Lambda }\right)
^{2\Delta _{{\rm s}}}(1-\bar{u})\;,  \label{G1}
\end{equation}
where $g_{1}\simeq 0.62$ is a universal constant \cite{AT:92}.

In the approximation of an infinite cutoff $\Lambda \rightarrow \infty$,
which physically means neglecting the discrete structure of matter, $\bar{u}
= u_{0}c_{t}^{2}/(u^{\ast }\Lambda a_{0}^{2})\rightarrow 0$ and the two
crossover parameters $\bar{u}$ and $\Lambda $ in the crossover equations
collapse into a single one, $\bar{u}\Lambda $, which is related to the
Ginzburg number $G$ by \cite{AT:92}
\begin{equation}
G=g_{0}\frac{(\bar{u}\Lambda )^{2}}{c_{t}}=g_{0}\frac{u_{0}^{2}v_{0}^{2}}{%
(u^{\ast })^{2}a_{0}^{4}\bar{\xi}_{0}^{6}} \;,  \label{Gi}
\end{equation}
where $g_{0}\simeq 0.028$ is a universal constant \cite{AT:92} and $\bar{\xi}%
_{0}=v_{0}^{1/3}c_{t}^{-1/2}=(c_{0}/a_{0})^{1/2}$ is the mean-field
amplitude of the power law for the correlation length. Note that the
Ginzburg number does not depend explicitly on the cutoff $\Lambda $ or on $%
\bar{u}$. This single-parameter crossover, i.e., the crossover for $\bar{u}%
=0 $, is universal and is indicated in Fig.~1 by a dashed-dotted curve. This
simplified description of the crossover is equivalent to the results of
Bagnuls and Bervillier \cite{BB:85} and of Belyakov and Kiselev \cite{BK:92}.

In the simulations \cite{LB:99}, each spin interacts equally with its $z$
neighbors lying within a distance $R_{{\rm m}}$ on a three-dimensional cubic
lattice. The effective range of interaction $R$ is then defined as $%
R^{2}=z^{-1}\mathop{\textstyle\sum}_{j\neq i}|{\bf r}_{i}-{\bf r}_{j}|^{2}$
with $|{\bf r}_{i}-{\bf r}_{j}|\leq R_{{\rm m}}$ \cite{LB:96}. We have
approximated the relation between $R$ and $R_{{\rm m}}$ by $R^{2}=\frac{3}{5}%
R_{{\rm m}}^{2}(1+\frac{2}{3}R_{{\rm m}}^{-2})$, as indicated in the insert
in Fig.~3. In order to compare the numerical results to the theoretical
prediction Eq.~(\ref{chi}), we need the range dependence of the parameters $%
c_{t}$ and $\bar{u}$. Indeed, the asymptotic $R$ dependence of $\bar{u}$
follows directly from simple scaling arguments~\cite{LB:96}, $\bar{u}=\bar{u}%
_{0}R^{-4}$, and $c_{t}$ varies as its square root, $c_{t}=c_{t0}R^{-2}$.
For a three-dimensional simple cubic lattice, $\Lambda =\pi $ \cite
{LF:90,F:74}, and we obtain for the Ginzburg number
\begin{equation}
G=G_{0}R^{-6}=0.28(\bar{u}_{0}^{2}/c_{t0}^{4})c_{t}^{3}=0.28(\bar{u}%
_{0}^{2}/c_{t0})R^{-6}\;.  \label{eq:ginzburg}
\end{equation}
The nonuniversal parameters $c_{t0}$ and~$\bar{u}_{0}$ have to be determined
from a least-squares fit to the numerical data for $\gamma _{{\rm eff}}^{+}$%
, which yielded $c_{t0}=1.72$ and $\bar{u}_{0}=1.22$ and hence $G_{0}\approx
0.24$. The solid lines in Fig.~1 indicate the corresponding theoretical
curves. It should be noted that these curves are calculated for each value
of $R_{m}$ separately; the piecewise continuous character of this
description directly reflects the fact that the crossover cannot be
described by a universal single-parameter function. Indeed, Fig.~1 also
shows two attempts to describe the data in terms of such a function. The
dash-dotted line corresponds to the limit $\bar{u}$ $\rightarrow 0$, whereas
the dotted curve corresponds to $\bar{u}_{0}=1.22$ and $\Lambda =\pi $ (a
continuation of the theoretical curve for $R=1$). We see that the actual
crossover lies between these two bounding curves, with $\bar{u}\simeq 0$ for
large $R$ and $\bar{u}\simeq 1.2$ for $R=1$. Thus, it is clearly seen that
without including the $R$ dependence of $\bar{u}$ it is impossible to
describe data for short interaction ranges $R_{{\rm m}}^{2}\leq 5$. The
dependence of $\bar{u}$ on~$R$ is shown in Fig.~3. The two adjustable
parameters $c_{t0}$ and $\bar{u}_{0}$ are strongly correlated and if one of
them is fixed at a predicted value, the quality of the description remains
the same. We hence tried to fit the data while keeping $c_{t0}$ fixed at the
theoretically predicted value $c_{t0}=2d=6$ \cite{FB:67,CT}. In this case a
fit of the same quality is obtained with $\bar{u}_{0}=1.22$, provided that $%
\Lambda \simeq 2\pi $. The value of $G_{0}\approx 0.24$ then remains
unchanged.

To describe the data below the critical temperature, a connection between $M$
and $\tau $ in zero field is to be found from the condition $(\partial
\Delta \tilde{A}/\partial M)_{\tau }=0$. The relation between $M$ and $\tau $
appears to be implicit and $\chi $ as a function of $\tau $ cannot be
expressed in an explicit form either. Of course, the parameters $c_{t0}$ and
$\bar{u}_0$ should be the same as for $T>T_{{\rm c}}$ and we hence kept them
fixed at the above-mentioned values. However, the parameter $G_0$ appearing
in Eq.~(\ref{eq:ginzburg}) will take a different value. We took this into
account by introducing a factor $G_{0}^{+}/G_{0}^{-}$ into the temperature
scale: $t\rightarrow t\cdot (G_{0}^{+}/G_{0}^{-})$. Figure~2 shows the
results for $T < T_{{\rm c}}$, where the factor $G_{0}^{+}/G_{0}^{-}$ was
included as an adjustable parameter. Our estimate $G_{0}^{-}/G_{0}^{+}=2.58$
must be compared with the theoretical result $G_{0}^{-}/G_{0}^{+}=3.125$~
\cite{LF:94}. Interestingly, $\gamma_{{\rm eff}}^{-}$ clearly shows a
minimum around $|t|R^{6}\sim 10^{2}$. This corroborates the nonmonotonic
character of the crossover of $\gamma _{{\rm eff}}^{-}$, earlier observed
for the two-dimensional Ising lattice~\cite{LB:97}, where the effect is much
more pronounced. We note that already in Ref.~\cite{BB:87} a field-theoretic
calculation of the crossover in the low-temperature regime has been given
(in the limit $\bar{u} \to 0$), but only recently this has been extended to
cover the full crossover region~\cite{PR:98}. Actually, also here a
nonmonotonicity in $\gamma_{{\rm eff}}^{-}$ has been observed.

In summary, we remark that although in general the theory contains two
crossover parameters $\bar{u}$ and $\Lambda $, only one parameter ($\bar{u}$%
) changes with the range of interaction. However, this does not mean that
the crossover is a universal function of $tR^{6}$. Indeed, the effective
range of interaction $R$ affects the behavior of $\gamma _{{\rm eff}}^{\pm }$
in a twofold way: through the Ginzburg number, which is proportional to $%
c_{t}^{3}$, and through the first Wegner correction, with an amplitude $%
\Gamma _{1}$ that is proportional to $(1-\bar{u})$ [Eq.\ (\ref{G1})]. Hence,
there is no way to describe the data for short interaction ranges without
allowing for $\bar{u}$ to become larger than unity and correspondingly $%
\Gamma _{1}$ to change its sign between $R_{{\rm m}}=2$ and $R_{{\rm m}}=1$
as indicated in Fig.~3. In previous publications we have shown that Eq.\ (%
\ref{chi}), derived from renormalization-group matching, gives an excellent
representation of the experimentally observed crossover behavior in simple
and complex fluids \cite{AS:95,MH:97,JS:98}. From the evidence presented in
this paper, we conclude that the same crossover model also yields a
quantitative description of the crossover critical behavior of a
three-dimensional Ising lattice.

\bigskip

{\bf Acknowledgments}

We acknowledge valuable discussions with M. E. Fisher and assistance from A.
A. Povodyrev. The research at the University of Maryland was supported by
DOE Grant No. DE-FG02-95ER-14509. E. Luijten acknowledges the HLRZ
J\"{u}lich for computing resources on a Cray-T3E.

\newpage

\begin{figure}
\centering\epsfclipon\epsfxsize 95mm
\leavevmode
\rotate[r]{\epsfbox{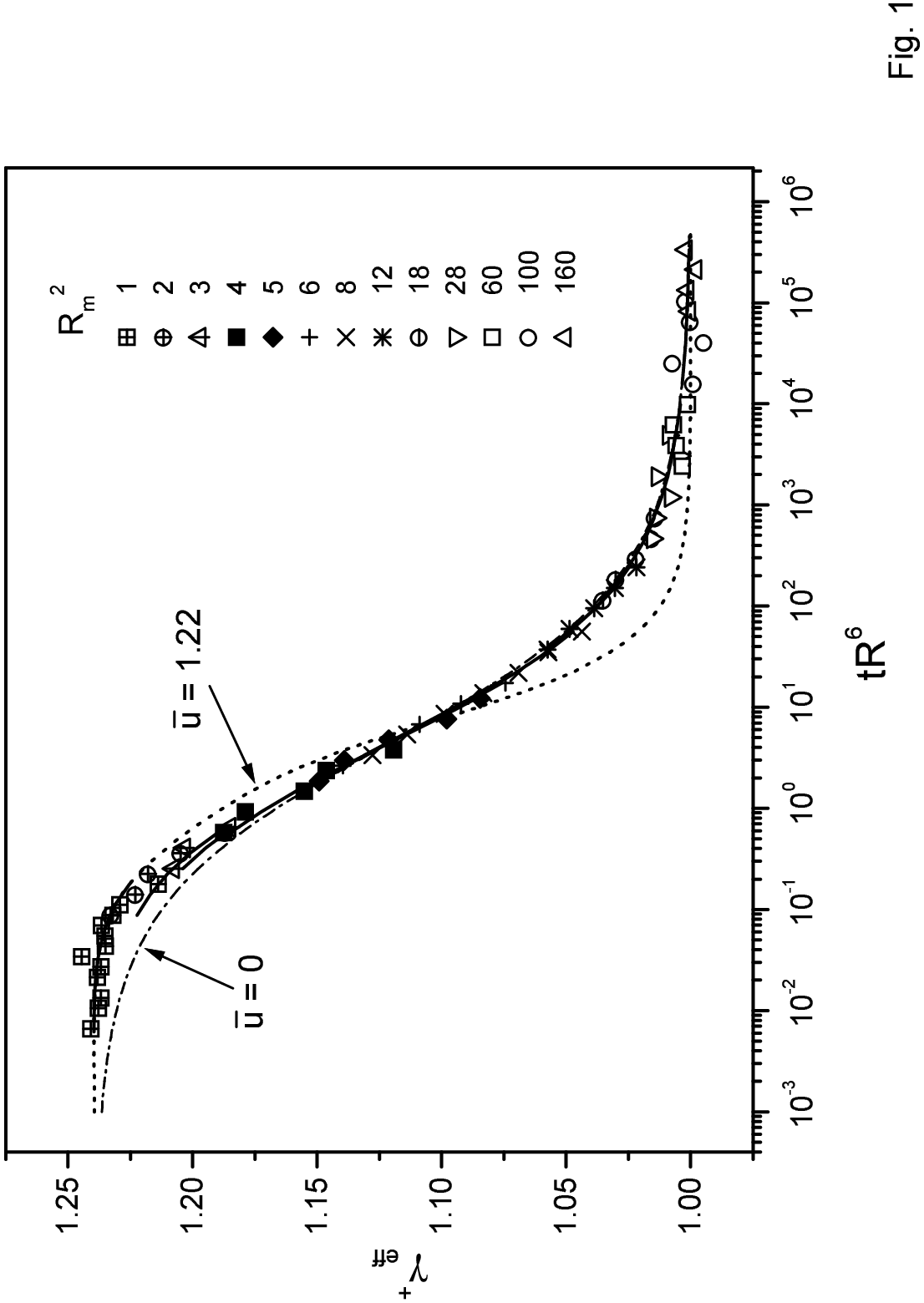}}
\caption[]{The effective susceptibility exponent $\gamma _{{\rm eff}}^{+}$
above $T_{{\rm c}}$. The symbols indicate numerical simulation data \cite
{LB:99}. The solid curves represent values calculated from Eq.\ (\ref{chi}).
The dashed-dotted curve corresponds to the limit $\bar{u}\rightarrow 0$. The
dotted curve is a continuation of the crossover curve for $\bar{u}=1.22$.
For clarity, the error bars have been omitted; they are all of the order of~$%
0.004$.}
\end{figure}

\begin{figure}
\centering\epsfclipon\epsfxsize 100mm
\leavevmode
\rotate[r]{\epsfbox{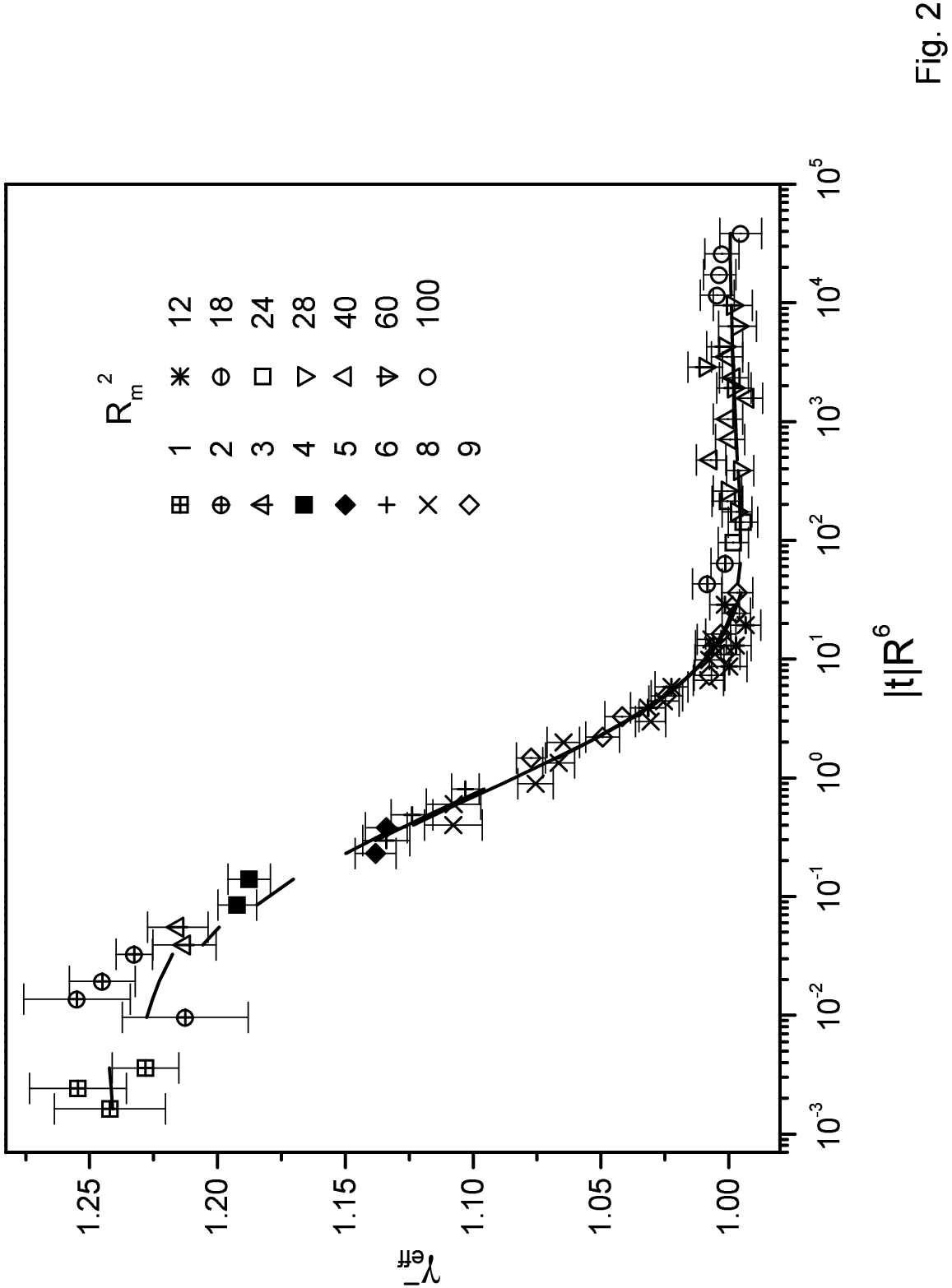}}
\caption[]{The effective susceptibility exponent $\gamma _{{\rm eff}}^{-}$
below $T_{{\rm c}}$. The symbols indicate numerical simulation data \cite
{LB:99}.  The solid curves represent values calculated from the
renormalization-group matching crossover model.}
\end{figure}

\begin{figure}
\centering\epsfclipon\epsfxsize 88mm
\leavevmode
\rotate[r]{\epsfbox{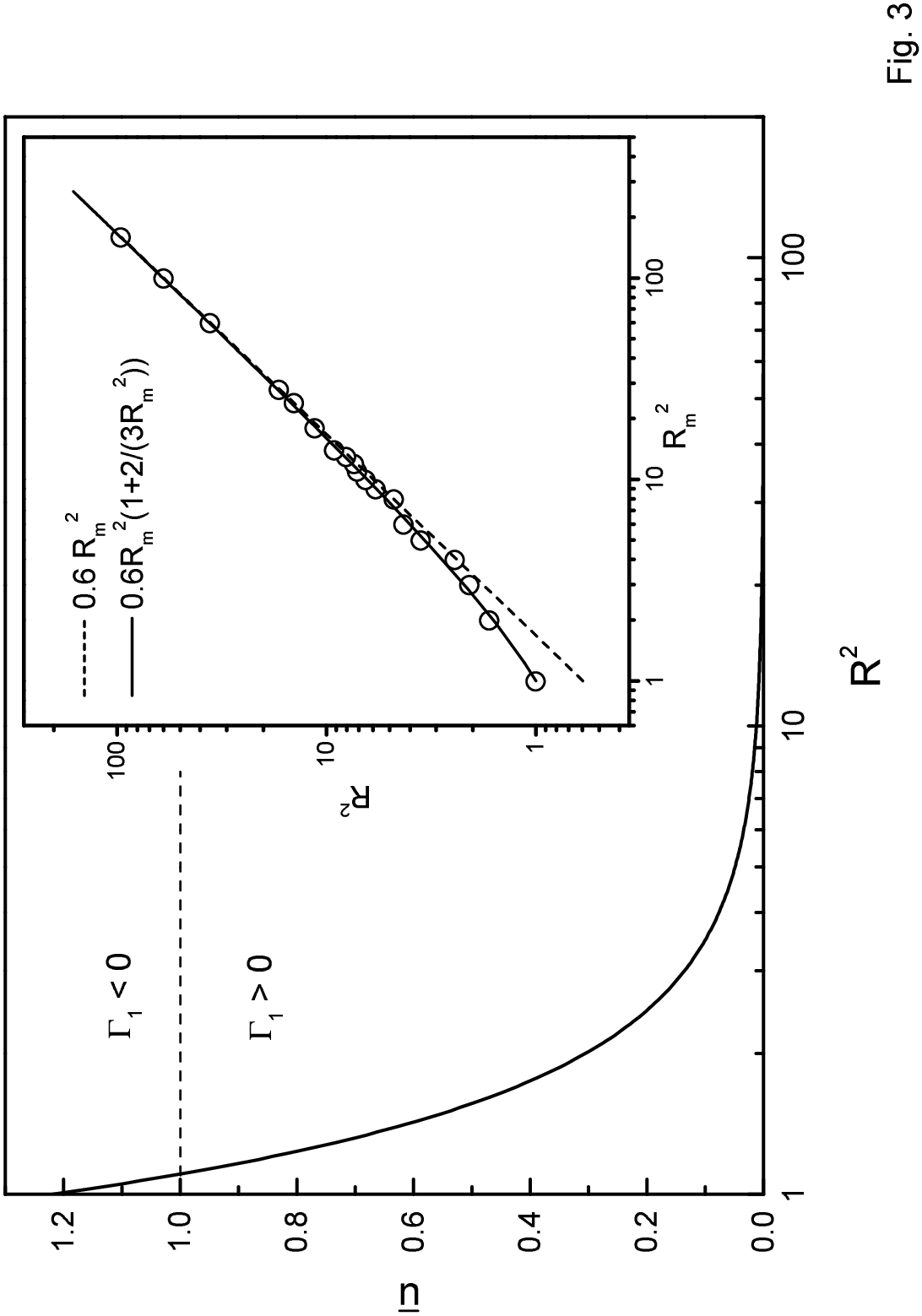}}
\caption[]{Dependence of the normalized coupling constant $\bar{u}$ on the
normalized interaction range $R$. Note that $\bar{u}$ becomes larger than unity
for very short interaction ranges. Insert: Effective range of interaction $R$
(open circles) plotted as a function of $R_{{\rm m}}$. The solid line
corresponds to the approximation mentioned in the text and the dashed line
represents the asymptotic behavior for large~$R$.}
\end{figure}

\end{document}